\documentclass[twocolumn, pra, amssymb]{revtex4}
\def\th{^{\mbox{\scriptsize th}}}
\def\c{\mathbb{C}}
\def\id{\mathbb{I}}
\def\vs{\vspace{.2cm}}
\def\o{\!\otimes\!}
\newcommand{\ket}[1]{\mbox{$| #1 \rangle$}}
\newcommand{\bra}[1]{\mbox{$\langle #1 |$}}
\newcommand{\proj}[1]{\ket{#1}\!\bra{#1}}
\def\tr{\mbox{tr}}
\def\r{\mbox{range}}
\def\k{\mbox{rank}}
\def\d{\mbox{dim}}
\def\s{\mbox{span}}
\def\h{\mathcal{H}}

\def\v{\ket{v}}
\def\pnv{\ket{v}\!_1\!\bra{v}}
\def\nv{\mathbf{0}}

\def\A{A_n(a|x)}

\begin{document}

\title{Extremal quantum correlations for $N$ parties\\
with two dichotomic observables per site}
\author{Llu\'{\i}s Masanes}
\affiliation{School of Mathematics, University of Bristol, Bristol
BS8 1TW, U.K.}
\date{\today}

\begin{abstract}
Consider a scenario where $N$ separated quantum systems are
measured, each with one among two possible dichotomic observables.
Assume that the $N$ events corresponding to the choice and
performance of the measurement in each site are space-like
separated. In the present paper, the correlations among the
measurement outcomes that arise in this scenario are analyzed. It
is shown that all extreme points of this convex set are attainable
by measuring $N$-qubit pure-states with projective observables.
This result allows the possibility of using known algorithms in
order decide whether some correlations are achievable within
quantum mechanics or not. It is also proven that if an $N$-partite
state $\rho$ violates a given Bell inequality, then, $\rho$ can be
transformed by stochastic local operations into an $N$-qubit state
that violates the same Bell inequality by an equal or larger
amount.
\end{abstract}
\maketitle

\section{Introduction and results}

A privileged method for contrasting quantum and classical physics
is by comparing the correlations among space-like separated events
that each theory predicts. This is so because one can find
constrains on the correlations predicted by each theory which are
independent of any model for the experiment. For instance, Bell
inequalities \cite{Be} are constrains on the correlations that
emerge from any possible experiment described by classical
physics. Analogously, the quantum Bell-type inequalities \cite{KT}
are constrains on the probability distributions generated by
measuring quantum systems, whatever the kind of systems and
measurements involved.\vs

Let us specify the scenario and the notation. Consider $N$
separated parties, denoted by $n=1,\ldots N$, each having a
physical system which can be measured with one among $M$
observables with $K$ outcomes each. The $n\th$ party observables
and outcomes are respectively denoted by $x_n\in \{1,\ldots M\}$
and $a_n\in \{1,\ldots K\}$. All the experimental information is
contained in the joint probability distribution for the outcomes
conditioned on the chosen observables $P(a_1\ldots a_N|x_1\ldots
x_N)$.\vs

The distributions predictable by local classical theories are the
ones that can be written as
\begin{equation}\label{CC}
    P(a_1\ldots a_N|x_1\ldots x_N) =
    \sum_\lambda\, p(\lambda) \prod_{n=1}^N P_n(a_n|x_n \lambda)
    \ ,
\end{equation}
where each $P_n(a_n|x_n \lambda)$ is a local distribution for the
outcome $a_n$ conditioned on the choice of observable $x_n$ and
the (classical) information shared among the parties $\lambda$
\cite{WWB}. Fixed $(N,M,K)$ to some finite values, the set of
distributions $P$ that can be written as (\ref{CC}) is a convex
polytope, which can be characterized by a finite set of linear
inequalities, the so called Bell inequalities. For a complete
introduction to Bell inequalities and related topics see
\cite{WWB}. However, only for some values of $(N,M,K)$ the Bell
inequalities are known. But this is not a grave problem, because
testing whether a given distribution $P$ can be written as
(\ref{CC}) or not, is a linear programming feasibility problem,
which for a wide range of $(N,M,K)$ can be solved easily. This is
not that simple in the case of quantum correlations, where the
problem is in general unbounded. In this paper we bound this
problem for the case $M=K=2$ and arbitrary $N$.\vs

Let us characterize the set of distributions that can be generated
within quantum theory. Suppose the $n\th$ party has a system with
Hilbert space $\h_n$, which is measured with the $M$ generalized
measurements $\{\A:\ a=1,\ldots K\}$ for $x=1,\ldots M$. These
POVMs satisfy $\A\geq 0$ for $a=1,\ldots K$ and $\sum_{a=1}^K \A
=\id_n$, for $x=1,\ldots M$ and $n=1,\ldots N$, where $\id_n$ is
the identity matrix acting on $\h_n$. For an introduction to the
formalism of generalized measurements see \cite{NC}. The
distributions predictable by quantum theory are the ones that can
be written as
\begin{equation}\label{QC}
    P(a_1\ldots a_N|x_1\ldots x_N) =
    \tr\!\left[\rho\, \bigotimes_{n=1}^N A_n(a_n|x_n)
    \right],
\end{equation}
where $\rho$ is a positive semidefinite matrix acting on
$\h=\bigotimes_{n=1}^N \h_n$ with $\tr\rho=1$. Fixed $(N,M,K)$ to
some finite values, the set of distributions $P$ that can be
written as (\ref{QC}) is convex, but not a polytope \cite{KT}.
These sets could also be characterized by Bell-type nonlinear
inequalities, but little is known about them \cite{T,M}. However,
if the dimension of the local Hilbert spaces $\h_n$ are fixed to a
finite number, deciding whether a given distribution $P$ can be
written as (\ref{QC}) or not (up to a chosen precision) is an
algorithmic task \cite{GLS}. Unfortunately, it is not known how to
bound the dimension of the local Hilbert spaces given $(N,M,K)$.
In this paper it is shown that for the case $M=K=2$ the extreme
points of the set (\ref{QC}) are attainable with $\h_n=\c^2$. This
allows for using the algorithms of \cite{GLS} in order to decide
whether a given distribution $P$ is quantum or not. \vs

From a fundamental point of view, it is also interesting to have a
finite characterization for quantum correlations. Actually, this
problem is proposed in the web page {\em ``Some Open Problems in
Quantum Information Theory" (problem 26.A)} \cite{KW}. In
particular, they rise the question whether the minimal dimension
sufficient to generate all quantum correlations for a given
$(N,M,K)$ is $K$. Here, this question is answered for the case
$M=K=2$ and arbitrary $N$. \vs

In the dichotomic case ($K=2$), one can reduce the amount of
experimental data by considering full-correlation functions
\begin{eqnarray}\label{cf}
    && C(x_1\ldots x_N)=\\ \nonumber
    && \sum_{a_1=1}^2 \ldots \sum_{a_N =1}^2
    (-1)^{\Sigma_{n=1}^N a_n}
    P(a_1\ldots a_N|x_1\ldots x_N)\ ,
\end{eqnarray}
instead of all the information $P(a_1\ldots a_N|x_1\ldots x_N)$.
That is, for each experimental setting $(x_1\ldots x_N)$ all the
information is summarized in the single number $C(x_1\ldots x_N)$.
Notice that in the general case $2^N-1$ numbers are necessary. In
\cite{WW}, the set of extremal quantum full-correlation functions
(\ref{cf}) is obtained for the case $M=K=2$. Here, the extremal
points are obtained for the general case, where all experimental
data is considered. \vs

As a corollary of the results proven in this paper, the following
is shown. If an $N$-partite state $\rho$ violates a given Bell
inequality (in the setting $M=K=2$), then $\rho$ can be
transformed by stochastic local operations into an $N$-qubit state
$\tilde{\rho}$ which violates the same Bell inequality by an equal
or larger amount. Here, by stochastic it is meant that the
operation can fail with some probability.

\section{Extremal quantum correlations for $M=K=2$}

In this section we prove the main results of the paper. Here and
in the rest of this document only the case $M=K=2$ is considered.

\subsection{Projective measurements are enough}

{\bf Lemma 1.} In the case $K=M=2$, all extreme points are
attainable by measuring pure states with orthogonal observables.
\vs

{\em Proof.} By the linearity of (\ref{QC}) with respect to $\rho$
it is clear that all extreme points can be expressed with $\rho$
being pure. The next holds for each party. A nice fact about
dichotomic POVMs $A(1|x)+A(2|x)=\id$ is that both operators can be
diagonalized in the same basis. Suppose that $\v$ is a
simultaneous eigenvector of $A(1|1)$ and $A(2|1)$, that is $A(a|1)
\v=\lambda_a \v$ for $a=1,2$, and $\lambda_1 +\lambda_2=1$. Then
we can write $A(a|1) =\lambda_a \proj{v} + \tilde{A}(a|1)$, where
$\tilde{A}(a|1)$ is a positive operator orthogonal to $\proj{v}$,
for $a=1,2$. It is clear that the POVM $\{A(1|1), A(2|1)\}$ can be
written as a convex combination of the two POVMs
$\{\proj{v}+\tilde{A}(1|1), \tilde{A}(2|1) \}$ and
$\{\tilde{A}(1|1), \proj{v}+ \tilde{A}(2|1)\}$ with the respective
weights $\lambda_1$ and $\lambda_2$. Continuing this procedure
with the rest of simultaneous eigenvectors one can express the
POVM $\{A(1|1), A(2|1)\}$ as a convex combination of projective
POVMs. $\Box$

\subsection{Non-factorizable extreme points}

Some extreme distributions have the property that a party can be
factorized, for instance
\begin{equation}\label{factorizable}
    P_1(a_1|x_1)P(a_2\ldots a_N|x_2\ldots x_N)\ .
\end{equation}
We are not interested in such extreme points, because they reduce
to the case of $N-1$ parties, which is already considered in the
$N$-partite case. We say that an extreme point is {\em
non-factorizable} if it cannot be written like
(\ref{factorizable}), for any of the parties. Notice that a
non-factorizable extreme point can be factorized in groups
containing more than one party.\vs

{\bf Lemma 2.} All non-factorizable extreme points are achieved
with observables $\{A_n(a|x)\}$ such that, every non-zero vector
$\v\!_n \in\h_n$ belongs to at most one of the four subspaces
$\{\r A_n(a|x): a,x=~1,2\}$, for $n=1,\ldots N$.\vs

{\em Remark.} This implies that all vectors in the range of
$A(1|1)$ have nonzero overlap with both, $A(1|2)$ and $A(2|2)$.
Loosely speaking, the two observables $A(a|1)$ and $A(a|2)$ ``do
not commute for each possible direction''.\vs

{\em Proof.} Suppose that a distribution $P$ is obtained by
measuring $\rho$ with the observables $\{A_n(a|x)\}$. Let us
consider the first party $n=1$. By Lemma 1 we assume that the four
operators $A_1(a|x)$ are projectors, hence their ranges are the
subspaces spanned by their corresponding eigenvectors with unit
eigenvalue. By orthogonality, no single non-zero vector belongs to
both, $\r A_1(1|x)$ and $\r A_1(2|x)$. Suppose that there is a
non-zero vector $\v\!_1\in~\h_1$ that belongs to $\r A_1(1|1)$ and
$\r A_1(1|2)$. Then we can write the four projectors as
\begin{equation}\label{}
    A_1(a|x)= \delta_{a,1}\pnv +\tilde{A}_1(a|x)\ ,
\end{equation}
where each $\tilde{A}_1(a|x)$ is a projector orthogonal to $\pnv$.
Let us define the probability $\pi=\tr[\pnv \rho]$ and the
normalized states
\begin{eqnarray}
    \rho_v &=& \frac{1}{\pi} \langle v| \rho\v\!_1\ ,\\
    \tilde{\rho} &=& \frac{1}{1-\pi} (\id_1-\pnv)\, \rho\,
    (\id_1-\pnv)\ ,
\end{eqnarray}
acting respectively on $\bigotimes_{n=2}^N\h_n$ and
$\bigotimes_{n=1}^N\h_n$.  Clearly, the original correlations
---for instance (\ref{QC})--- can be expressed as the mixture
\begin{eqnarray}
    &&\tr\!\left[\rho\, \bigotimes_{n=1}^N A_n(a_n|x_n) \right]
    \\
    = \pi\, \delta_{a,1}\!\!\!\!
    &&\tr\!\left[\rho_v \bigotimes_{n=2}^N A_n(a_n|x_n)\right]
    \nonumber\\
    + (1-\pi)\!\!\!\!
    &&\tr\!\left[\tilde{\rho}\, \tilde{A}_1(a|x)
    \bigotimes_{n=2}^N A_n(a_n|x_n)\right]
    \label{mixture}\ .\quad
\end{eqnarray}
The first term in the right-hand side is factorizable, hence, we
ignore it. In the second term, neither the matrix $\tilde{\rho}$
nor the operators $\tilde{A}_1(a|x)$ have any overlap with
$\v\!_1$. Now, relabel $\tilde{\rho} \rightarrow \rho$ and
$\tilde{A}_1(a|x) \rightarrow A_1(a|x)$, and consider the second
term in the right-hand side of (\ref{mixture}). We can repeat the
process until no single vector in $\r A_1(1|1)$ is contained in
$\r A_1(1|2)$. The same can be done to the other three pairs of
operators: $\{A_1(1|1),A_1(2|2)\}$, $\{A_1(2|1),A_1(1|2)\}$,
$\{A_1(2|1),A_1(2|2)\}$, and also to the rest of parties
$n=2,\ldots N$. If the initial correlations $P$ are a
non-factorizable extreme point, after all this procedure, we
obtain an extreme point with the property stated in Lemma 2.
$\Box$\vs

{\bf Lemma 3.} All non-factorizable extreme points are attainable
with a state acting on $\bigotimes_{n=1}^N\h_n$, where every local
Hilbert space $\h_n$ has even dimension, and $\k A_n(a|x)=\d
\h_n/2$ for $a,x=1,2$ and $n=1,\ldots N$.\vs

{\em Proof.} The following analysis can be applied to every party,
and we omit the subindex $n$. Suppose that $\{\ket{u_1},\ldots
\ket{u_r}\}$ is an orthonormal basis for the subspace $\r A(1|1)$,
where $r=\k A(1|1)$. Because the direct sum of $\r A(1|2)$ plus
$\r A(2|2)$ is the local Hilbert space $\h_n$, every vector in
this basis can be expressed as a direct sum $\ket{u_i}=\ket{u_i^1}
+ \ket{u_i^2}$, where $\ket{u_i^a}=A(a|2)\ket{u_i} \in\r A(a|2)$
for $i=1,\ldots r$. According to Lemma 2, both $\ket{u_i^1}$ and
$\ket{u_i^2}$ are not null, otherwise $\ket{u_i}$ would belong to
$\r A(1|2)$ or $\r A(2|2)$. If $\d[\s\{\ket{u_1^1},\ldots
\ket{u_r^1}\}] <r$ there exists a set of coefficients
$\{c_1,\ldots c_r\}$, not all being zero, such that $\sum_{i=1}^r
c_i \ket{u_i^1} =\nv$. This implies that $\sum_{i=1}^r c_i
\ket{u_i} = \sum_{i=1}^r c_i \ket{u_i^2}$, and consequently that
$\sum_{i=1}^r c_i \ket{u_i}$ belongs to $\r A(1|1)$ and $\r
A(2|2)$, against Lemma 2. Therefore, it must be the case that
$\d[\s\{\ket{u_1^1},\ldots \ket{u_r^1}\}] =r$. This implies that
$\k A(1|1) \leq \k A(1|2)$, but applying the same argument from
the point of view of $A(1|2)$ we obtain $\k A(1|2) \leq \k
A(1|1)$, so both ranks are equal. One can repeat this argument for
the pairs of projectors $\{A(1|1), A(2|2)\}$ and $\{A(2|1),
A(1|2)\}$, concluding that $\k A(a|x)=r$ for $a,x=1,2$. We finish
the proof by noticing that, by construction $\d\h_n= 2r$, which is
an even number. $\Box$

\subsection{Qubtis are enough}

The main result of the paper is the following\vs

{\bf Theorem 4:} In the case $K=M=2$, all quantum extreme points
(\ref{QC}) are achievable by measuring $N$-qubit pure states with
projective observables.\vs

{\em Proof:} Suppose that the distribution $P(a_1\ldots
a_N|x_1\ldots x_N)$ is obtained by measuring $\rho$ with the set
of observables $\{A_n(a|x)\}$. Here we assume that the observables
$A_n(a|x)$ are of the form specified in Lemma 1, Lemma 2 and Lemma
3. The following analysis can be applied to every party, and thus,
we omit the subindex $n$. Define $r=\k A(1|1)$ and the two
matrices $G_a=A(1|1) A(a|2) A(1|1)$ for $a=1,2$. Because $A(1|1)$
is the identity in the subspace $\r A(1|1)$ and $G_0 + G_1 =
A(1|1)$, there exists a simultaneous eigenbasis for $G_0$ and
$G_1$ in the subspace $\r A(1|1)$, denoted by $\{\ket{v_1},\ldots
\ket{v_r}\}$. Define the pair of vectors $\ket{v_k^a} = A(a|2)
\ket{v_k}$ and the two-dimensional subspace $E^k=\s\{\ket{v_k^1},
\ket{v_k^2}\}$ for $k=1,\ldots r$. Because $\ket{v_k} \propto
G_a\ket{v_k}= A(1|1) A(a|2) A(1|1) \ket{v_k}$ then $\ket{v_k}
\propto A(1|1) \ket{v_k^a}$, which implies that for any $\v\in
E^k$ we have $A(1|1) \v\propto \ket{v_k}$. Denote by
$\ket{v_k^\bot}$ any non-zero vector in $E^k$ orthogonal to
$\ket{v_k}$. Due to the above discussion %$\a\v\propto \ket{v_k}$
we have that $\ket{v_k^\bot}=[A(1|1) + A(2|1)] \ket{v_k^\bot}=
A(2|1) \ket{v_k^\bot}$, and then $\ket{v_k^\bot} \in \r A(2|1)$.
Summarizing, the subspace $E^k$ contains one, and only one, vector
(up to a constant factor) from each of the four spaces $\r
A(1|1)$, $\r A(2|1)$, $\r A(1|2)$, $\r A(2|2)$. These vectors are
respectively $\ket{v_k}, \ket{v_k^\bot}, \ket{v_k^1},
\ket{v_k^2}$.

In each of the subspaces $E^k$ we define the pair of projective
measurements
\begin{eqnarray}
    A^k(1|1) = \frac{\proj{v_k}}{\langle v_k | v_k\rangle}
    && A^k(2|1) = \frac{\proj{v_k^\bot}}{\langle v_k^\bot | v_k^\bot \rangle}
    \ ,\quad\quad\\
    A^k(1|2) = \frac{\proj{v_k^1}}{\langle v_k^1 | v_k^1 \rangle}
    && A^k(2|2) = \frac{\proj{v_k^2}}{\langle v_k^2 | v_k^2 \rangle}
    \ .\quad\quad
\end{eqnarray}
Suppose that the $N$ parties have made this procedure. That is,
from $n=1,\ldots N$, the $n\th$ party has the $r_n$ bidimensional
subspaces $E^k_n$ and pairs of observables $A^k_n(a|x)$ for
$k=1,\ldots r_n$. We also denote by $E^k_n$ the projector onto the
subspace $E^k_n$. Define the probability distribution
\begin{equation}
    \pi^{[k_1\cdots k_N]} =
    \tr[\rho\, E^{k_1}_1 \o\cdots\o E^{k_N}_N]\ ,
\end{equation}
and the normalized $N$-qubit states
\begin{equation}\label{}
    \rho^{[k_1\cdots k_N]} =
    \frac{E^{k_1}_1 \o\cdots\o E^{k_N}_N\, \rho\,
    E^{k_1}_1 \o\cdots\o E^{k_N}_N}
    {\pi^{[k_1\cdots k_N]}}\ ,
\end{equation}
for $k_n=1,\ldots r_n$ and $n=1,\ldots N$. Due to the fact that
for each party $n$, the subspaces $E_n^1,\ldots E_n^{r_n}$ are
mutually orthogonal and add up to the whole local Hilbert space
$\h_n = \bigoplus_{k=1}^{r_n} E_n^k$, we can conclude the
following. The original distribution $P(a_1\ldots a_N|x_1\ldots
x_N)$ can be written as the mixture
\begin{eqnarray}\label{average}
    \!\!\!\!&& P(a_1\ldots a_N|x_1\ldots x_N) = \\ \nonumber
    \!\!\!\!&& \sum_{k_1=1}^{r_1}\cdots\sum_{k_N=1}^{r_N}
    \pi^{[k_1\cdots k_N]}\,
    P^{[k_1\cdots k_N]}(a_1\ldots a_N|x_1\ldots x_N)\ ,
\end{eqnarray}
in terms of the more extreme distributions
\begin{eqnarray}\label{Nq}
    P^{[k_1\cdots k_N]}(a_1\ldots a_N|x_1\ldots x_N)
    \\ \nonumber
    =\tr\!\left[\rho^{[k_1\cdots k_N]}
    \bigotimes_{n=1}^N A^{k_n}_n(a_n|x_n) \right]\ .
\end{eqnarray}
Each distribution $P^{[k_1\cdots k_N]}$ is obtained by measuring
an $N$-qubit state with projective observables. Concluding, if the
original distribution $P(a_1\ldots a_N|x_1\ldots x_N)$ is an
extreme point, it can be obtained by measuring an $N$-qubit state
with projective observables. $\Box$

\section{Violation of Bell inequalities after LOCC}

In this section we derive a corollary that follows from the
previous results. Given an $N$-partite state $\rho$, consider the
$N$-qubit states $\tilde{\rho}$ that can be obtained from $\rho$
with some probability, when the parties perform protocols
consisting of local operations and classical communication (LOCC).
We do not care about the success probability as long as it is
nonzero. This set of transformations is called stochastic-LOCC or
SLOCC.\vs

Consider the Bell inequality specified by the coefficients
$\{\beta(a_1\ldots a_N|x_1\ldots x_N)\}$. That is, all
distributions of the form (\ref{CC}) satisfy
\begin{eqnarray}\label{BI}
    \sum_{a_1, x_1 =1}^2 \cdots \sum_{a_N, x_N =1}^2
    \beta(a_1\ldots a_N|x_1\ldots x_N)
    \times \\ \nonumber \times
    P(a_1\ldots a_N|x_1\ldots x_N)
    \geq 0\ .
\end{eqnarray}
And suppose that the $N$-partite state $\rho$ violates this
inequality when measured with the observables $\{A_n(a_n|x_n)\}$
\begin{eqnarray}\label{BI}
    \sum_{a_1 x_1 =1}^2 \cdots \sum_{a_N x_N =1}^2
    \beta(a_1\ldots a_N|x_1\ldots x_N)
    \times \\ \nonumber \times
    \tr\!\left[\rho\, \bigotimes_{n=1}^N A_n(a_n|x_n) \right]
    < 0\ .
\end{eqnarray}
Let us apply the methods used in the proofs of Lemma 2 and Theorem
4 to prove the following result.\vs

{\bf Corollary 5.} If an $N$-partite state $\rho$ violates the
Bell inequality given by the coefficients $\{\beta(a_1\ldots
a_N|x_1\ldots x_N)\}$, then $\rho$ can be transformed by SLOCC
into an $N$-qubit state $\tilde{\rho}$ that violates the
inequality $\{\beta(a_1\ldots a_N|x_1\ldots x_N)\}$ by an equal or
larger amount.\vs

{\em Proof.} To prove this statement, let us show how to construct
a rank-two projector for each party ($X_n$ for $n=1,\ldots N$),
such that the $N$-qubit state
\begin{equation}\label{pro}
    \tilde{\rho}= \frac{X_1\otimes\cdots\otimes X_n\, \rho\,
    X_1\otimes\cdots\otimes X_n}
    {\tr\left[\rho\, X_1\otimes\cdots\otimes X_n\right]}\ ,
\end{equation}
violates the Bell inequality $\beta(a_1\ldots a_N|x_1\ldots x_N)$
by an equal or larger amount.

In the proof of Lemma 2, each party keeps removing a particular
kind of vectors $\v\!_n$ from its Hilbert space $\h_n$. At the end
of this procedure the final Hilbert space $\h'_n$ is a subspace of
$\h_n$. The projection of the observables $\{A_n(a|x)\}$ onto
$\h'_n$ satisfy the properties stated in Lemma 2. The rank-two
projector $X_n$ has support on $\h'_n$, and is specified in the
next. In the proof of Theorem 4 we define the family of $N$-qubit
states $\rho^{[k_1\cdots k_N]}$. Each of them is obtainable from
$\rho$ by performing the SLOCC transformation (\ref{pro}) with
projectors $X_n=E_n^{k_n}$. By equation (\ref{average}), it is
clear that if $P(a_1\ldots a_N|x_1\ldots x_N)$ violates a Bell
inequality, there must exist one distribution $P^{[k_1\cdots k_N]}
(a_1\ldots a_N|x_1\ldots x_N)$ which also does. The corresponding
$N$-qubit state $\tilde{\rho}= \rho^{[k_1\cdots k_N]}$ is what we
where looking for. By convexity, the violation attained by
$\tilde{\rho}$ is never smaller than the one by $\rho$. $\Box$\vs

For some authors, Bell inequalities need not to be facets of the
classical polytope (\ref{CC}). For them, any linear inequality
satisfied by all distributions of the form (\ref{CC}) is a Bell
inequality. Remarkably, Corollary 5 also holds for these more
general definition of Bell inequalities.

\section{Conclusions}

We have considered a Bell-experiment scenario for $N$ parties,
each with two dichotomic observables. The correlations that arise
when measuring quantum systems in such scenario form a convex set.
We have proven that all the extreme points of this set are
achievable by measuring $N$-qubit pure states with projective
observables. This answers the question risen in Problem 26.A of
\cite{KW} for the case $M=K=2$. It would be very interesting to
prove that the minimal local dimension sufficient for generating
all the extreme points of the quantum set $(N,M,K)$ is always $K$.
Unfortunately, the techniques used in our proofs are not directly
applicable for larger values of $M$ or $K$. \vs

More practically, the obtained characterization allows for an
algorithmic procedure to decide whether a particular distribution
$P$ is predictable by quantum mechanics or not \cite{GLS}. \vs

We have also shown that if a state $\rho$ violates a given Bell
inequality, then $\rho$ can be transformed by stochastic local
operations into an $N$-qubit state which violates the same Bell
inequality by an equal or larger amount. This result has
interesting consequences when considering the violation of Bell
inequalities after LOCC, and in the regime where a large number of
copies of the state ($\rho^{\otimes n}$) are jointly measured.
This will be investigated in a forthcoming paper.

\acknowledgements

The author is thankful to Antonio Ac\'{\i}n and Andreas Winter for
discussions. This work has been supported by the U.K. EPSRC's
``IRC QIP''.


\begin{thebibliography}{99}

\bibitem{Be} J. S. Bell; Physics {\bf 1}, 195 (1964).

\bibitem{KT} L. A. Khalfin, B. S. Tsirelson; Found. Phys. {\bf
22}, 879 (1992).

\bibitem{WWB} R. Werner, M. M. Wolf; (Review article for the QIC journal)
quant-ph/0107093.

\bibitem{NC} M. A. Nielsen, I. L. Chuang; {\em Quantum computation and
    quantum information} (Cambridge University Press, Cambridge, 2000).

\bibitem{T} B. S. Tsirel'son; J. Sov. Math. {\bf 36}, 557 (1987).

\bibitem{M} Ll. Masanes; quant-ph/0309137.

\bibitem{GLS} M. Gr\"otschel, L. Lovasz, G. Schrijver;
{\em Geometric Algorithms and Combinatorical Optimization}
(Springer-Verlag, Berlin 1988).

\bibitem{KW} O. Krueger, R. F. Werner; quant-ph/0504166. www.imaph.tu-bs.de/qi/problems/

\bibitem{WW} R. F. Werner, M. M. Wolf; Phys. Rev. A. {\bf 64},
032112 (2001).

\end{thebibliography}
\end{document}